# Novel Concept of Circular-Linear Energy Recovery Accelerator to Probe the Energy Frontier

I. V. Konoplev*[1,4], S.A. Bogacz[2], Ya. Shashkov[3], M.A. Gusarova[3]

[1] *JAI, Department of Physics, University of Oxford, Oxford, OX1 3RH*

[2] *Thomas Jefferson National Accelerator Facility, Newport News, Virginia 23606, USA*

[3] *National Research Nuclear University MEPhI, Moscow, Russia*

[4] *Sevastopol State University, Sevastopol, Russia*

**Abstract.**

Energy-frontier accelerators provide powerful tools performing high precision measurements confirming the fundamental of the physics and broadening new research horizons. Such machines are either driven by circular or linear accelerators. The circular machines, having the center-mass (CM) energy values reaching 200 GeV (for leptons) and above, experience beam energy loss and quality dilution, for example, due to synchrotron radiation, limiting the overall CM energy achievable and requiring a constant energy top-up to compensate the loss and the beam quality dilution. Linear colliders overcome these limitations, while the finite capabilities of generating high average current beams limits the luminosity. This is partially compensated by the quality of the colliding beams. In this work, we suggest a novel design of circular-linear accelerator based on the merging of the "non-emitting", low-energy storage rings and energy recovery linear accelerators. We suggest using the recently considered dual-axis asymmetric cavities to enable the operation of such a system, and in particular the energy recovery from spent, high-intensity beams. The machine considered, under the scope of the SNOWMASS-2021 initiative, can be potentially used to reach ultimate energy frontiers in high-energy physics as well as to drive next generation light sources. The merging of circular and linear systems, and applications of dual axes cavities, should allow the maintaining of high beam quality, high luminosity, and high energy efficiency - 'The Best of Both Words'. It also offers a flexible energy management, opening clear opportunity for reducing the running cost. We note that the numbers shown in the paper are for illustration purpose and can be improved further.

* corresponding author, e-mail address: i.konoplev202@gmail.com

1. **Introduction**

Energy-frontier particle accelerators are among the most exciting, complex, challenging, and expensive research instruments [1-6]. They are required to drive a broad range of fundamental physics research [1-6], and can also generate the brightest light, in order to carry out studies in life sciences, create new materials, and learn about matter under the extreme conditions [7-10]. The highest energies machines, from multi-GeV to several TeV, (ILC, FCC, CLIC) [1-6, 11-13] are to probe the most fundamentals concepts of physics and to search for the basic building blocks of matter. Currently, the most accepted centre-mass operating points are around 91 GeV (Z pole), 160 GeV (W± pair-production threshold), 240 GeV (ZH production) and 340–365 GeV (tt threshold and above), and there are plans to build machines in the TeV range for further studies. Conventional machine design is normally based on a single-axis RF structures (both normal and superconducting), but often fall short of satisfying all critical requirements simultaneously (i.e. beam quality and intensity, energy efficiency, operational cost, and compactness). Most of the accepted state-of-the-art designs [1-17], while reaching the energies required, have very large footprint and show the same signs of limitations and drawbacks as described above. Energy-frontier accelerators which use conventional techniques are rapidly approaching the limits of feasibility, and novel accelerator technologies, such as wakefield acceleration, are not yet mature enough to replace conventional accelerators. To resolve this and avoid stagnation, new ideas are required to revitalize conventional RF accelerators, making their construction feasible and run-costs non-prohibitive. This paper is will discuss one such idea. It is based on preliminary studies, which have been recently carried out by different groups. The numbers shown are for illustration purpose and authors believe they can be significantly improved during the appropriate following studies.

The use of the energy recovery linacs (ERL) for particle colliders was suggested in 1965 [15] to overcome the limitations associated with energy efficiency. Application of the ERL concept appears frequently in the literature [14-20], and a recent design [14], is based on circular collider with energy recover capabilities and damping rings. This conceptual design [14] is based on a 100 km FCC tunnel with two ERLs (each 33.7 GeV SRF linacs) separated by 1/6th of the 100 km circumference to boost and recover the energy. To reach the energy required, the charged particle bunches have to make 4 full circles, and after the interaction point they would undergo energy recovery making another 4 full loops. This eight pass system also has two 1 GeV "cooling" rings, which can be used for the bunch storage.

As indicated in the paper [14], due to the use of the ring there, energy is still lost on both the accelerating and decelerating passes. Though synchrotron radiation (SR) emitted power is significantly reduced (as compared with the baseline FCC), it is calculated that the loss due to synchrotron radiation should be around 30 MW per beam. The advantage and attractiveness of this design is due to the possibility to employ energy recovery as well as to accumulate and condition the electron/positron bunches in the rings.

There are also challenges associated with increasing the luminosity of the system. To increase the luminosity, one would face similar limitations as in circular machines. Increasing the average current would lead to the possibility of the excitation and development of the beam break-up instabilities (BBU) [21-24]. We also note, that the development of the beam break-up instabilities is a significant shortcoming of the conventional ERLs, and has hampered the development of the high-intensity applications based on ERLs, including colliders and high-power light sources.

This work will describe opportunities and concepts, which have risen from the recent developments of the dual-axis asymmetric cavities for energy recovery linacs (ERL) [25-31], to overcome some of the limitations associated with BBU instabilities, improve energy consumption and energy management flexibility. In this paper we suggest a novel concept which is based on merging the advantages of the circular (capability to accumulate beams) and linear (capability to deliver good quality beams) accelerators. It is based on application of low-energy, "non-emitting" (1 GeV) large (1 km radius) storage rings and two dual-axis energy recovery linacs. The aim of the design is to realise a system capable of accumulating charged particle beams without energy or quality loss, condition the beams, boost these high intensity beams to the energy frontiers while preserving the beam quality, recuperate the energy of the particle beams, and return them to the storage rings. The conceptual design presented should enable "flexible energy management" (without beam loss) of the whole system, as well as a beam which can be used for high-energy physics and a driver for the next generation of light sources. The differences between this work and previous conceptual designs are in the dual-axis geometry of the systems, including, collider and injection/beam-dumping recirculating energy recovery system (RERS), use of the non-emitting storage rings (enabling use of the bunches to drive light sources), and broad application of the dual axis asymmetric cavities for energy recovery at all stages of particle acceleration/deceleration (A/D).

In this work, the fundamental challenges which conventional accelerators are facing will be discussed in Sections 2. In Section 3 the preliminary design (schematics) of the sections of the system will be shown and a specific examples using data adopted from the conventional systems will be discussed. We will also discuss the properties of dual-axis accelerating structures. Section 4 will conclude with a summary of the concept and discuss the steps required to bring this novel concept to fruition.

## 2. Limitation and challenges.

Particle accelerators which are capable of reaching the TeV range of energies are generally used to collide the particles in order to study the products of the beam-beam or beam-target impacts. The defining properties of these colliders are the center of mass (CM) energies of the particles, as well as the luminosity (i.e. a measure of how many collisions are taking place between the colliding beams in a collision area per second (units of cm$^{-2}$s$^{-1}$)). It is desirable to have a high luminosity, as this increases the number of the collisions and hence the number of the events that can be registered and analysed accurately. In general, the luminosity of a beam-beam collider can be defined, assuming a collision between two (transverse) Gaussian bunches, as:

$$L = \frac{N_1 N_2 f_c}{4\pi \sigma_x \sigma_y} \eta(\phi_{cr}, A) \qquad (1a)$$

where $N_{1,2}$ is the number of particle in each bunch, $\sigma_{x,y} = \sqrt{\varepsilon_{x,y} \beta_{x,y}}$ are the Gaussian bunch transverse dimensions (RMS), which correlated to the beam emittances $\varepsilon_{x,y}$ and the respective $\beta_{x,y}$ functions at the collision point, $f_c$ is the frequency of the collisions, and $\eta(\phi_c, A)$ is the efficiency of the collision, which depends on the crossing angle of the two beams, $\phi_{cr}$, and parameter $A$, which is proportional to the length of the bunch $\sigma_z$ and inversely proportional to the $\beta_{x,y}$ functions at the interaction point (IP). It is clear that $\eta(\phi_{cr}, A)$ can be controlled by beams optics, and for clarity, and only in this paper, we will assume that $\eta(\phi_{cr}, A) \cong 1$. For circular colliders, the luminosity is limited and scaled as the square of the synchrotron radiation power emitted by each circulating beam, $P_{SR,i} = eV_{SR}f_c N_i$. Assuming that two colliding beams have an identical number of particles (i.e. $N_1 = N_2$) and substituting this into expression (1a), the luminosity can be expressed more simply as

$$L = \frac{I^2}{4\pi e^2 f_c \sigma_x \sigma_y} , \qquad (1b)$$

where $I = \left(\frac{P_{SR}}{V_{SR}}\right)$ is the current parameter. Assuming a fixed $I$, and referring to (1b), increasing the luminosity in circular machines requires a decrease of $f_c$ and at least one of the transverse dimensions $\sigma_{x,y}$ (i.e. "flat" beams). In many cases, these parameters ($f_c$, $\sigma_{x,y}$) are optimised for the beam transport, limiting the flexibility of these parameters. It should also be noted that the beam energy loss, $\Delta E$, due to synchrotron emission scales with the beam energy as $\gamma^4$, where $\gamma$ is the relativistic Lorentz factor. This creates a significant challenge for circular machines, as $\Delta E$ is proportional to the radius ($\rho$) of curvature as $\rho^{-2}$.

Other limitations observed in TeV energy circular machines which are important to take into consideration, are associated with the beam quality degradation linked to a several phenomena, including natural energy spread and emittance dilution due to quantum excitations [16,31,32]. The undisputable advantage of the circular machines is the possibility to increase luminosity by continuously increasing the current parameter $I$. However, this would correspond to an increase of the beam energy loss due to the radiation $P_{SR,i}$. To mitigate these limitations, the circular machines are designed with large radii. For example, the circumference of the 350 GeV FCC (e$^+$e$^-$ machine) is around 100 km, limiting the loss $P_{SR,i}$ to 50MW per beam ( $P_{SR,i} \sim 1/\rho^2$).

Linear colliders are capable of avoiding many of the limitations of the circular machines. A number of linear colliders have been considered and studied, including SLAC, CLIC, and ILC [1,5, 33,34]. The most prominent designs, which have been developed in the last 20 years, are considered to be CLIC and ILC. These machines are TeV class e$^+$e$^-$ colliders based on either SRF standing wave (SW) cavities (ILC) or traveling wave (TW) room temperature accelerating structures (CLIC). The lengths of both colliders are comparable with the diameter of the FCC: ILC is 31 km (for the 1TeV design) and CLIC is 11 km (for the 380 GeV design) and upgradable to 50 km (for 3TeV). Like the FCC, the luminosity targets for these linear machines are expected to be at or above $1 \times 10^{34}$cm$^{-2}$s$^{-1}$. Although the linear colliders luminosity could be increased by optimising the beam parameters (such as $\sigma_{x,y}$), one of the challenges is to increase the repetition rate (i.e. increase the current parameter). For the ILC, it is projected that only some percentage of the accelerating buckets will be used, and the train of 1312 bunches will be generated with repetition rate 5Hz. The design uses 6.28 km circumference, 5 GeV cooling rings, and the spent bunches will be directed to the beam dump. The fact that the spent bunches are lost after a single interaction leads to a significant drop of the efficiency of such a machine. The energy efficiency and management are become the decisive criteria for both circular and linear colliders.

In general, the design of any of such a complex machine is based on set of compromises. At a relatively low energies (sub-TeV), a circular machine has a better ratio of the power (MW) required to the luminosity generated, making them more attractive as compared with the linear machines. At high energies (above TeV), there is no such advantage, and linear machines become more attractive. In this work, we suggest the following steps, which may improve the overall efficiency:

- Substituting the conventional SRF single axis cavitiy with dual axis structures, enabling the energy recovery of high current beams,
- Re-circulating and re-using the spent beam using non-radiative, multi-purpose storage rings.

3. **Preliminary concept of the Circular-Linear Energy Recovery Collider (C-LERC)**

To overcome the challenges discussed, we suggest Circular-Linear Energy Recovery System (CLER-S), based on dual-axis, asymmetric accelerating/decelerating (A/D) cavities. This system can be used for HEP to drive a dual axis collider (CLER-C), to drive next generation light sources (for example free-electron lasers (CLER-FEL)), or for dual purposes. As the most important part of the system discussed in this work is related to high energy colliders, the remainder of this text will refer to it as CLER-C.

The schematics of the novel conceptual design are shown in Figures 1a and 1b. In all the figures shown, the arrows indicate the beam travel directions. The only difference between Figures 1a and 1b is in position of the rings. Figure 1b is more complex, as the rings are moved to a single location. At this stage, we do not see obvious advantages/disadvantages for either scheme, and the justifications for the selection of a specific design would be a future development. All stages of the beam acceleration and deceleration (A/D) use the dual axis asymmetric cavities [28-31]. This type of the cavitiy is suggested due to their EM properties (i.e. good mode separation, localisation of the high order modes, and possibility to independently control the operating mode field amplitudes in A/D sections). These properties would potentially allow one to increase the currents at which BBU initiates, which would in turn allow the total beam power (current) to remain the same or increase. This would then increase the luminosity of the collider. Furthermore, the machine could operate using the spent beam, assuring good energy efficiency of the system.

In the subsections below, the concepts of each sub-system will be discussed, starting with the dual-axis asymmetric cavity. An example of the parameters of the CLER-C is presented in the Table 1. The set of parameters are based on the possibility of merging the properties of circular and linear colliders. We also suggest to use non-emitting storage rings as a drivers for FEL stations to broaden the application of the whole system and increase its efficiency and impact.

## 3.1 Dual axis SRF cavity

For clarity, it is assumed that the cavities operate at 1.3 GHz. Figures 2 and 3 illustrate possible designs of standing wave, dual-axis, asymmetric SRF cavities. The basic parameters for dual axis cavities are presented in Table 2. Unlike conventional (single axis) systems, the dual axis cavity allows for the separation of the accelerating and decelerating beam trajectories. This mitigates the BBU instability excitations by breaking the feedback between parasitic HOMs excited in one of the sections and beam dynamics in other section. It has been demonstrated [29-31] that the HOMs excited in a specific section of the cavity are localised, and will not impact the beam dynamics inside the other section. Figures 2 and 3 show the 2D schematics of the multi-cell cavities, which are composed of low-loss, Tesla-like cells and dual-axes bridge cells. The most complex part of the structure is the bridge cell, which has been recently constructed and tested at JLab [25-27]. The operating field distributions are shown in Figures 2 and 3, illustrating the possibility to vary the field amplitudes in such a structure by changing the cells geometries (Figs. 2b and 3a) and the number of the cells (Fig. 3b). Independent field control allows the spent beams to be used in the ERL. Changing the field amplitudes increases the acceptance of the decelerating buckets. To observe the efficient beam capturing threshold, the field amplitude ($E_{th}$) should be above $E_{th}^0 = \frac{E_0}{\gamma_0}$, where $E_0 = \frac{\pi m_e c^2}{\lambda e}$, $\gamma_0$ is the relativistic Lorentz factor of the unperturbed beam, $\lambda$ is the operating wavelength, $c$ is the speed of light, and $e$ is the electron charge. If the spent beam is not monoenergetic, and its energy is inside the interval $\gamma \in [\gamma_0 - \delta\gamma; \gamma_0]$, the amplitude of the field in decelerating section should be increased to $E_{th}^* \cong E_{th}^0(1 + \frac{\delta\gamma}{\gamma_0})$ to capture most of the beam. The possibility to increase the amplitude of the decelerating field may also compensate for some electron beam current loss.

## 3.2 Beam injection system.

The use of Recirculating Energy Recovery Systems (RERS) may be more energy and cost efficient, as compared with the use of a conventional ERL. We are considering RERS as "Plan A", though, using a large ERL injector will remain "Plan B", should the RERS prove infeasible. In Figure 4, the schematics of the initial stages of electron beam generation, energy boosting to 1 GeV, and its injection into storage ring are shown. Figure 4a illustrates the RERS, while Figure 4b shows ERL injector. In case of the RERS (Figure 4a), the beams are accelerated by 1.3 GHz SRF ERLs to the initial 40 MeV and injected into the recirculating system. Similar systems have recently been developed and their operation has been demonstrated [17,18]. The beams would undergo four stages of acceleration in four full loops. A single loop has a three 1.3 GHz, 80 MeV SRF energy recovery capable modules, allowing the beam to gain/recover 240 MeV (Figure 4), resulting in the 1 GeV/40 MeV beam respectively after four loops. The linear non-scaling fixed field alternating gradient optics [17] can be used to transport the beams with the dual energies (in our case the energy factor below 2) along the one set of arcs (right on Fig.4a). In the first arc, beams of energies 120 MeV and 200 MeV are expected, while in the second arc it is 360 MeV and 440 MeV, in the third arc it is 600 MeV and 680 MeV, and in the fourth it is 840 MeV and 920 MeV. Along the second set of arcs (left on Fig. 4a), the mono-energetic beams of energies 40 MeV, 280 MeV, 520 MeV, and 760 MeV will be transported. The spent beams are returned from RERS to the 40 MeV dual axis ERL to be decelerated and driven to the beam collector with energies below 1 MeV. The bunch parameters are expected to be within the following limits: single bunch charge from 0.5 nC to 1.5 nC, repetition rate from 300 MHz to 700 MHz, RMS bunch length from 0.5 mm to 2 mm, transverse *rms* bunch dimensions, $\sigma_{x,y}$, from 0.002 mm to 0.5 mm, average beam current from 0.1 A to 1 A. Before injecting the 1 GeV bunches into the cooling rings, they can be transported through a conditioning line (compression, cooling, etc.) to match the beam parameters to the beams inside the rings. In Table 3, the parameters of the RERS are presented.

In Figure 4b, an alternative way to inject the beam is shown. It illustrates the 1.3 GHz, dual axis asymmetric ERL. Considering an average accelerating field in the range from 15 MeV/m - 25 MeV/m, the length of such an ERL is expected to be up to a 100 m. The main advantage of such an ERL is in its relative simplicity, as no complex magnet system to guide beam is required. The possible drawback is an increased number of the 80

MeV accelerating modules from 3 to 12 (if one assumes that the beam after the photocathode has an energy around 40 MeV). This will increase the initial cost of the injection system, its footprint and possible energy, maintenance and cryogenic run costs. Clearly the final choice of the injection system should be made during the more detailed studies, analysis and design.

**3.3 Storage Ring**

The beams are accumulated in the storage rings (StRg) prior to injection. The bunches are accumulated, recirculated, and conditioned (if necessary) in such rings, and in Figure 1 the two large radius (1000 m), low energy (1 GeV) storage rings (red circles) are shown. The specific locations of the rings at this stage are not essential. We illustrate that the rings can be either at the ends of the Linear Energy Recovery Collider (LERC) as shown in Fig. 1a, or constructed in the same place in the middle of the system. The storage rings suggested are to be used for: collection and accumulation of bunches from both the injectors and the collider (re-circulated beams) conditioning of the bunches bringing them to the required specifications; re-injection for repeat application; re-direction for dumping.

To minimise the emission of the synchrotron radiation from the bunches inside the ring, we suggest accumulating the bunches at a relatively low energy (1 GeV) while maintaining the ring radius as large as possible. This would reduce the energy loss and beam quality degradation. These parameters are proposed as the beam energy loss due synchrotron radiation generation scales as $\gamma^4$ and $1/\rho^2$. Although similar rings had already been designed for ILC, we suggest reducing the beam energy from 5 GeV to 1 GeV to further reduce energy loss. The choice of 1 GeV is also dictated by the compromise between SR energy loss and the ability to use such beams for the photon generation without additional energy boosting. It is assumed that the rings will unable to maintain the beams for as long as required. Figure 5a shows the 1 GeV beams from the RERS are injected into the storage rings. There are two rings, each having a 1 km radius (6.3 km circumference) and enabling the storage of up to 1 A of average current, with the expected energy loss due to synchrotron radiation not exceeding 0.1 keV per turn. Accepting beams from the injector and recuperating beams from the linear collider should increase further the luminosity and the efficiency of the machine. The storage rings should also be used to control the quality of the bunches. If electron bunches cannot be conditioned, they are moved to the collector via deceleration sections of the ERLs (green ellipse) to be dumped with energies below 1 MeV after passing through the ERL-injector.

We also suggest using the rings as the photon factories to drive EUV, X-ray, and γ-photon radiation sources via diversions of some of the bunches into, for example, Free Electron Laser (FEL) stations as shown in Figures 5a and 5b. This will broaden the research horizons of the project and attract industrial partners to develop applications such as lithography below 5 nm. Figure 5b illustrates one of possible FEL stations installed in parallel to the main ring. In the case, some of the bunches can be re-directed toward the FEL stations and either returned or dumped after the use. Considering that the circumference of the ring suggested is 6.3 km, and assuming a typical undulator length of 150 m, one may expect a several FEL stations installed along (in parallel) the ring. Each FEL station could have energy boosting and recovering systems, which would allow the generation of photons in a wide range of the energies (from EUV, X-ray to γ-photons). One may also position the FEL stations inside the ring (Fig. 5a) along the diameter (to minimise the footprint), getting two contra-propagating beams of energies up to 10 GeV (each) and study, for example, Compton scattering of FEL photons on the 10 GeV electrons. Such configuration will also allow the use of the ERL dual axis system, which would minimise the energy cost. The use of dual axis A/D cavities will allow maintaining the efficiency of the whole ring system, while the presence of the several FEL stations may attract industrial partners, making the collider more cost efficient and generating stronger societal impact and support. One can consider using such a storage ring as a base for industrial FEL stations applicable, for example, to lithography below 5 nm. The FEL-stations would maximise the research efficiency and social impact of the whole facility.

## 3.4 Linear Energy Recovery Collider

To boost the beam energy, the beams are injected from the storage rings into the linear accelerators with energy recovery capabilities. The beams are accelerated toward each other along the accelerating arms of the cavities from the initial energy 1 GeV, up to the required energy, $W_{f,i}$. After the interaction point, which is crossing point similar to ILC and can be supported by similar optics as designed for the ILC, both beams are directed and transported through the decelerating arms of the dual-axis structures, reducing the beam energies to 1 GeV. Using the configuration described, i.e. storage ring (with SRF dual axis cavities) +SRF linac, the repetition rate of the bunches in LERC is only limited by the bunch repetition rate observed for non-emitting storage ring. This allows assuming that LERC will be able to operate either at high-repletion rate (similar to FCC). One notes, that the bunches will be injected into StRgs by RERS (operating like PERL in CW regime) indicating that in some configurations LERC can operate in CW regime as well.

In Figure 6 the Linear Energy Recovery Collider (LERC) is demonstrated. Once the beams are conditioned in the storage rings (StRg), they are injected into LERC. The LERC's left and right 10 km long energy recovery linacs will be comprised from dual-axis asymmetric 1.3 GHz SRF cavities (Fig. 2), operating at an average gradient of up to 25 MV/m. Similar to ILC, a 4.5 km long beam delivery system can be used to bring the two beams into collision with a 14 mrad crossing angle, at a single interaction point (as designed for ILC). Figure 6 shows the schematic of the LERC. Similar to ILC, the LERC consists of two 10 km ERLs (assuming modest 25 MeV/m constant average accelerating gradient observed in SRF cavity). To start, we suggest following the general system design chosen for ILC [33] to realize the CM energy of 500 GeV (head on collision) after the bunches accelerated along the first halves of the linacs. After the collision, the particles will undergo the deceleration to 1 GeV, and they will be re-injected into the storage ring for conditioning and re-use. It is possible that further upgrades can be achieved by either increasing the accelerating gradient of accelerating structures, while maintaining the same length of the tunnel or increasing the length of the tunnel at different stages of the collider development. An example of the beam parameters are shown in Table 3.

An alternative of such "straight forward" approach is adding "dog-bone" arcs, as discussed in [20, 32, 35] or using a structure similar to RERS. However, there are challenges and in spite the fact that the idea of the using the arc in a dog-bone configuration introduced for muon colliders in [35] is very fruitful, one has to remember that the muon mass nearly 200 times larger as compared with an electron or positron. This means that for the same arc used for muon, the amount of the beam energy loss due to synchrotron radiation will be $1.6*10^8$ times larger. Using the expression $P_{SR} = 0.089 \left[\frac{MW}{turn}\right] \frac{E_b^4(GeV) I_b(A)}{eR(m)}$, one can estimate a single bunch of current, $I_b$, energy loss, noting that for LEP [6] at energy of 100 GeV, 6 mA (per bunch) and 3 km radius, the total power loss was around 20MW or 3 GeV per turn. Taking this into account, the advantages of using the "dog-bone" configuration at the final stage, currently, seems to be unclear as compared with simple increase of the length of the LERC's linear accelerator parts.

## 4. Conclusion

We discussed the challenges which conventional accelerators are facing. Alternative solutions, as compared either with circular or linear designs, have been suggested. Using these solutions, the concept of energy frontier accelerators has been presented and discussed. The machine presented is a lepton-lepton collider, capable of operating at the energy frontiers while being energy efficient, and enabling flexible energy management. The machine merges several concepts, which have being recently developed and tested. It exploits energy recovery using dual-axis asymmetric SRF 1.3 GHz cavities, at all stages, and due to using the non-emitting storage-ring the linear collider based on SRF dual axis cavities may operate either in high repetition or CW regimes (depending on specific design). The collider design is made of several sections: a recirculating energy recovery system, which is used for initial injection of the 1 GeV electron beam; a low energy (1 GeV), large diameter storage ring to reduce energy losses due to coherent synchrotron radiation and allow storing the beams and their use while the collider is not in operation; and linear ERL boosters with "dog bone" arcs capable of reaching several energy bands and energy recuperation. The reasons to apply the dual axes asymmetric cavities are discussed. The structures of the cavities and the operating field profiles were shown, and the advantages of using these cavities have been demonstrated.

The possibility to use the storage rings as photon factories to boost the research and societal impact of the project and to attract possible industrial partners is discussed. The schematic diagrams of several FELs stations, which would allow to use the machine while the collider is idle, were shown. This may also make the project more attractive for countries to host it as it may to help boost research and development in other areas like nano-electronics, biology, etc.

## 5. Acknowledgement

One of the authors Prof I.V. Konoplev would like to thank Sevastopol State University for partial support of the work via internal grant (42-01-09/169/2021-3). Prof I. V. Konoplev would also like to thank STFC UK for partial support via grant (The John Adams Institute for Accelerator Science, ST/P002048/1). Work at Jefferson Lab has been supported by U.S. DOE under contracts DE-AC05-06OR23177 and DE-SC0012704. We would also like to acknowledge useful discussions with Dr. Ryan Bodenstein.

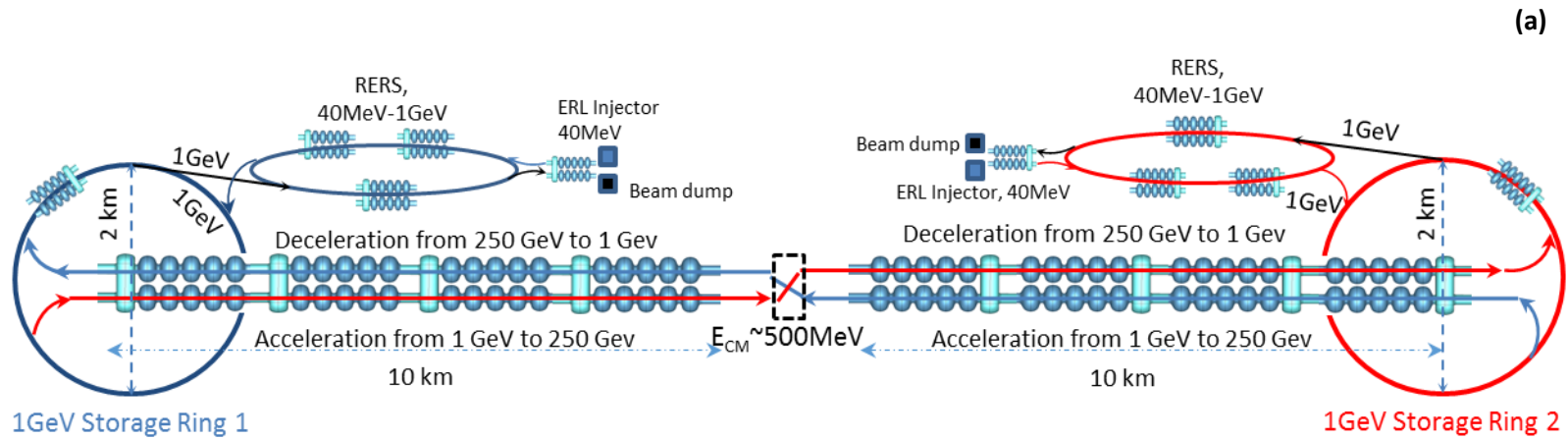

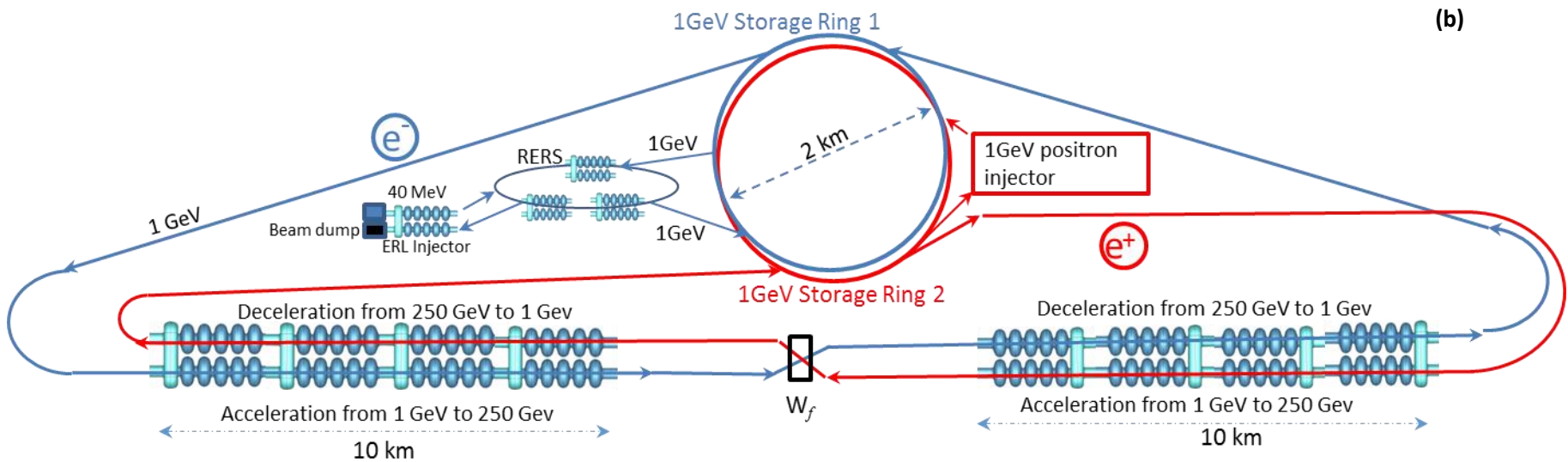

Fig.1

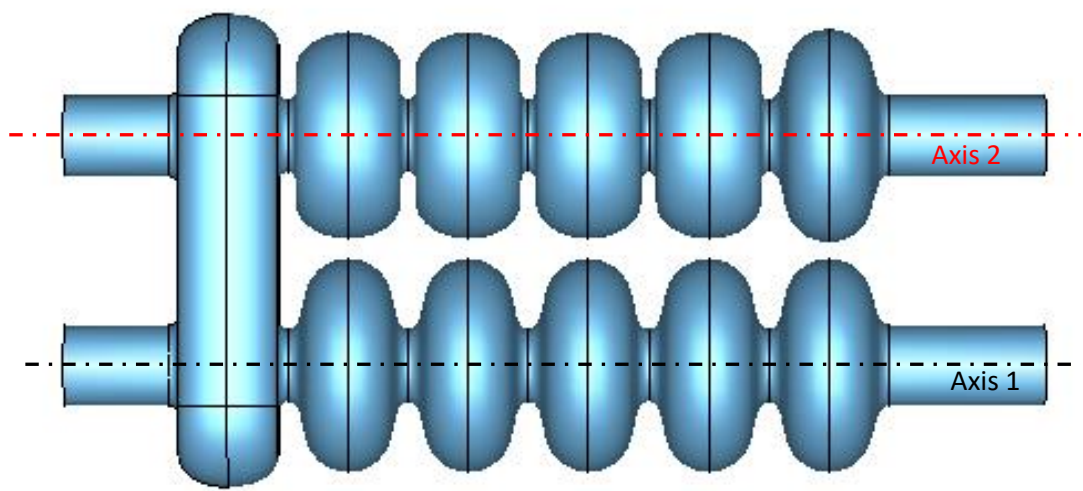

(a)

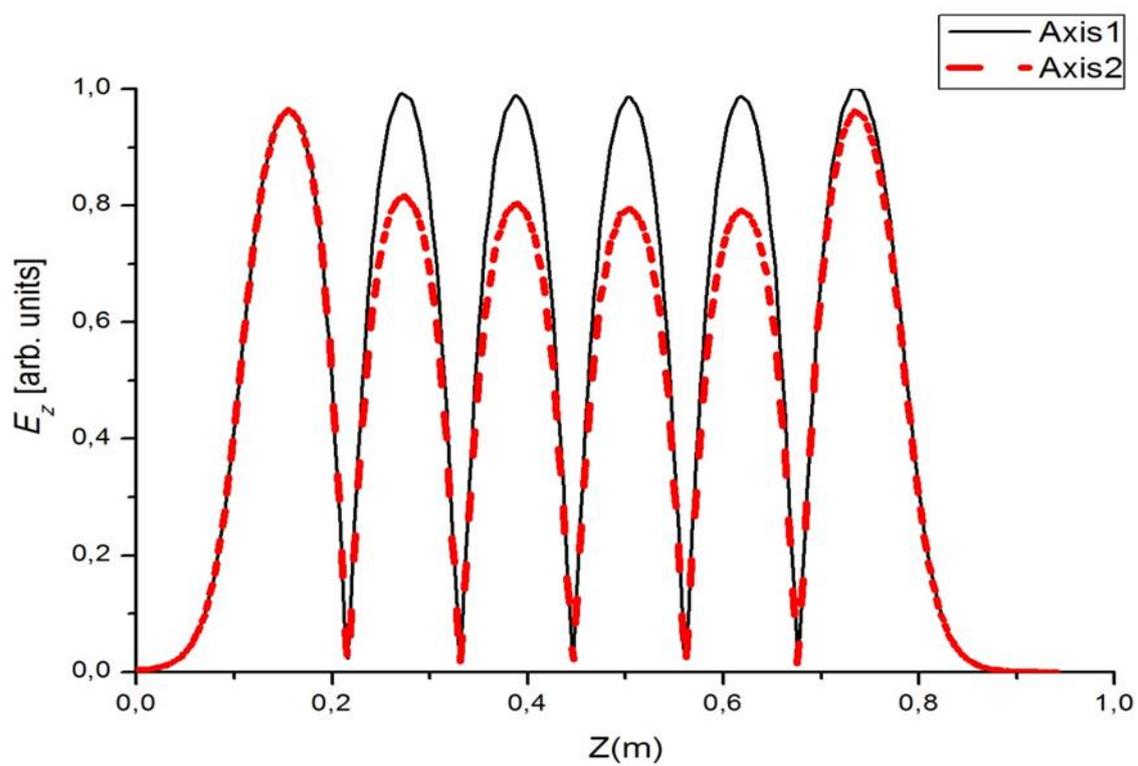

(b)

Fig.2

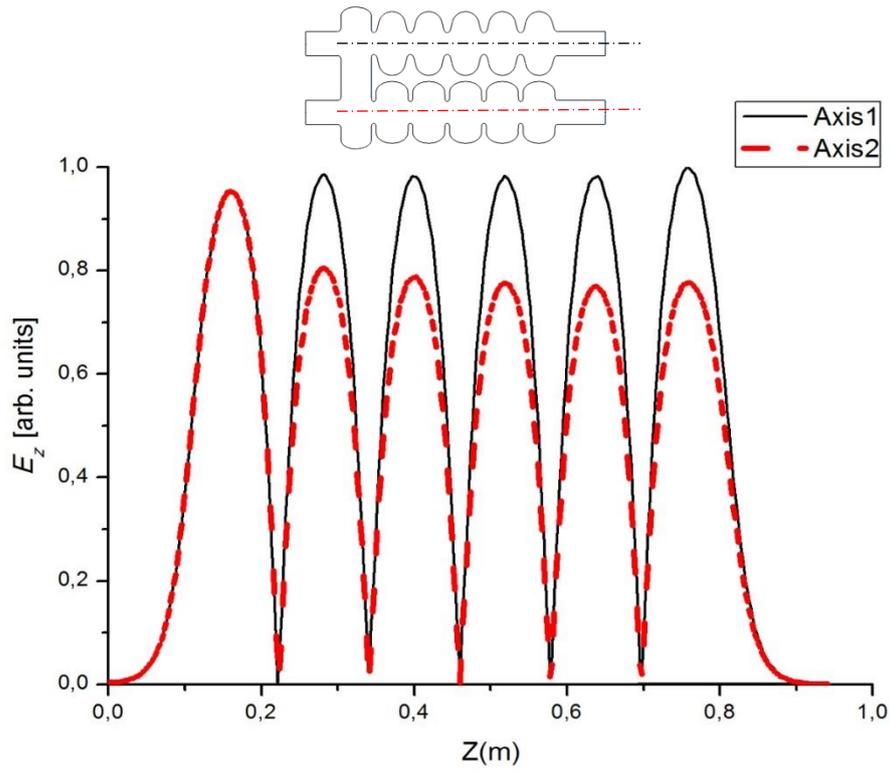

**(a)**

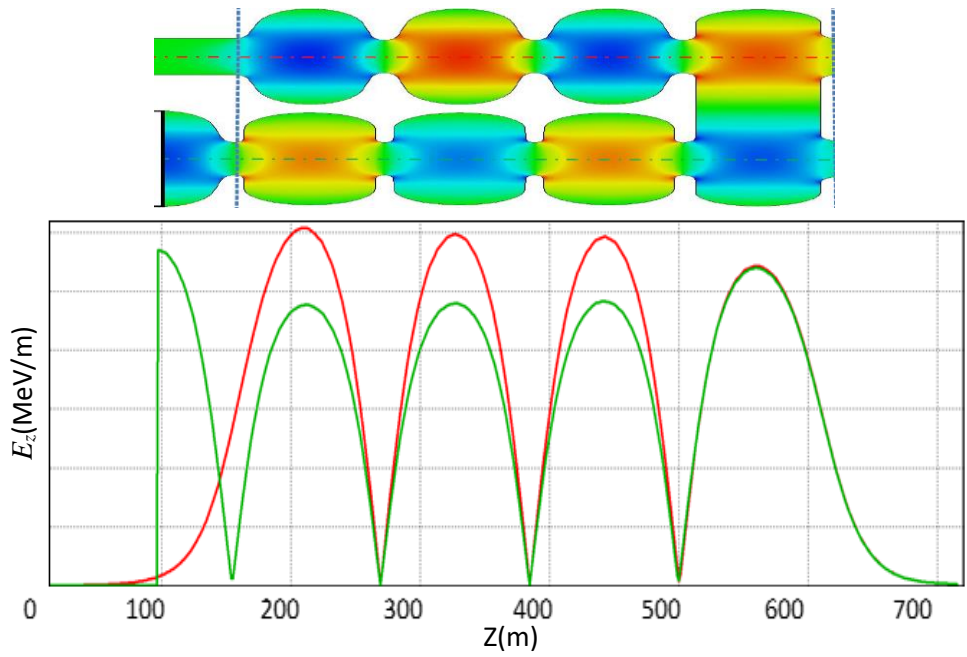

**(b)**

Fig.3

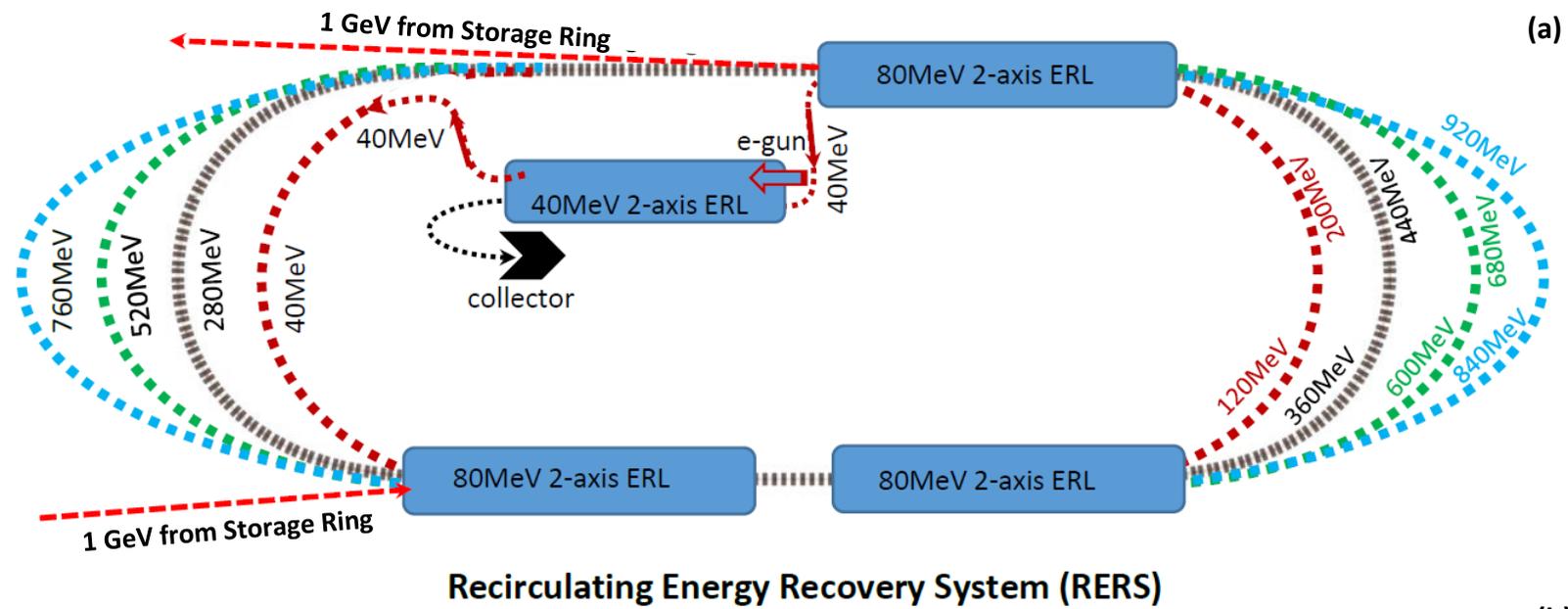

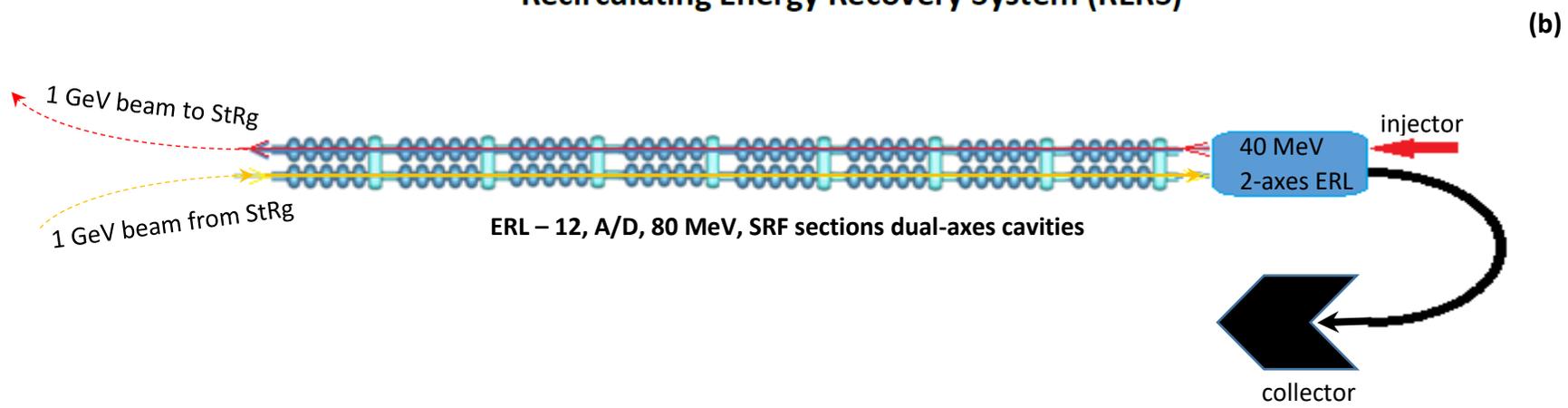

Fig.4

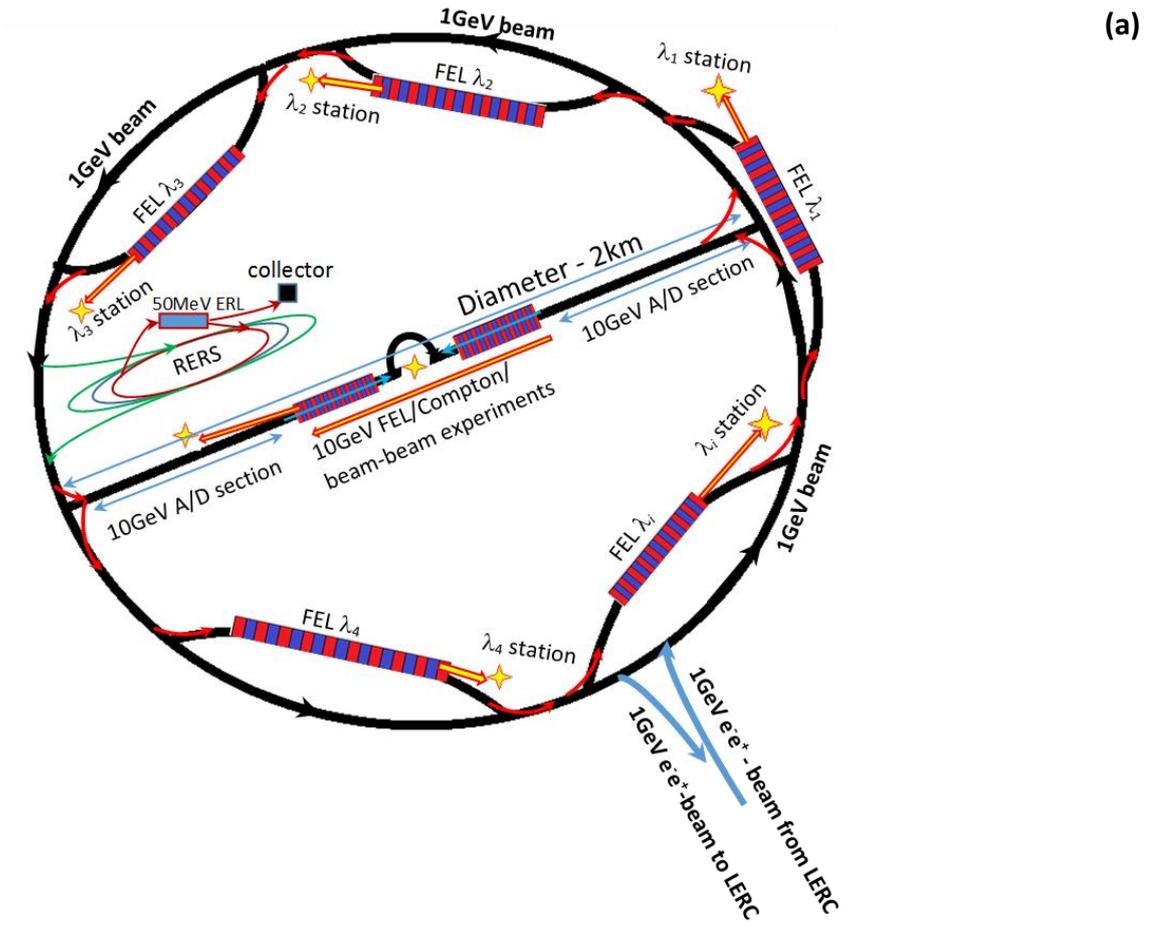

(a)

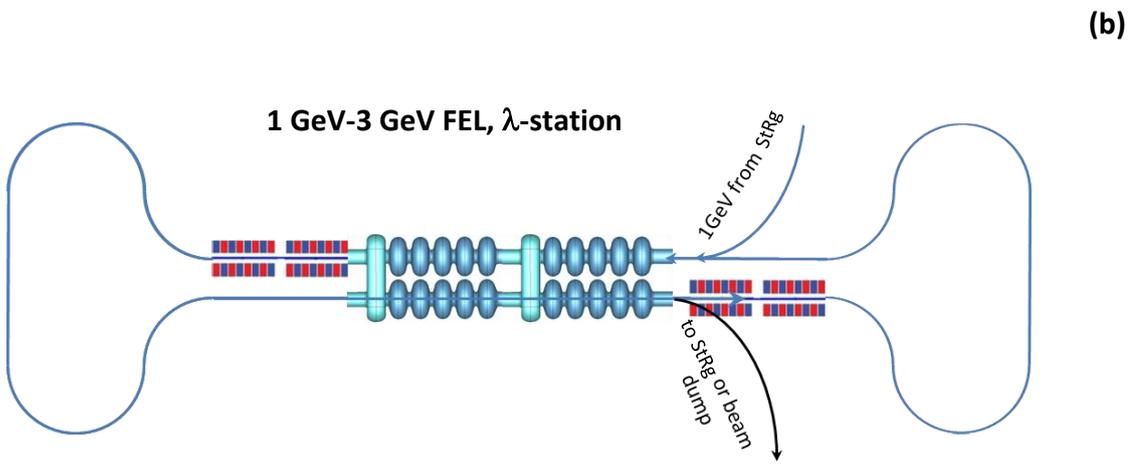

(b)

**1 GeV-3 GeV FEL, λ-station**

Fig.5

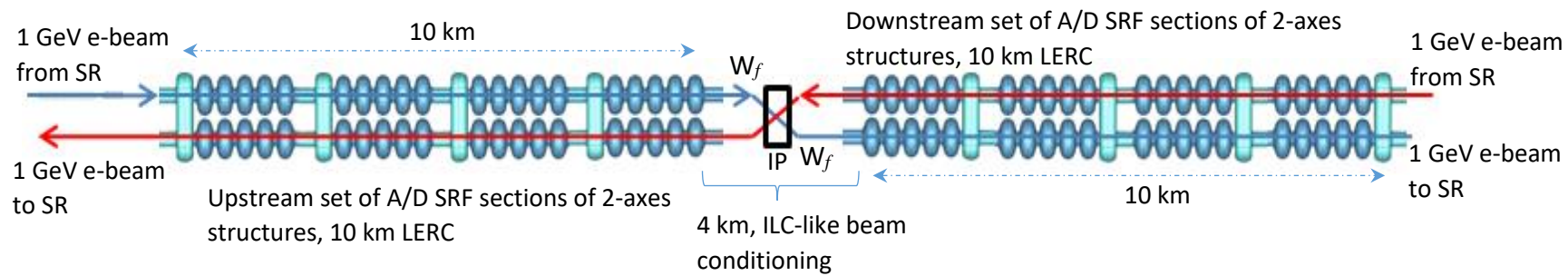

Fig.6

**Table 1. General outlook and comparison of the CLER-C with FCC and ILC**

|  | FCC- ee[hh] | ILC ee[ee] | CLER-C |
|---|---|---|---|
| **Type** | SRF Circular | SRF Linear | SRF ERL combine |
| **Main dimensions (km)** | 97.75 | 31 | 31 |
| **Energy (GeV)** | 120 | 250 (Base line) | 250 |
| **Peak Lum. ($10^{34} cm^{-2} s^{-1}$)** | 7 | 1.8 | >7 |
| **Interaction region** | 2(3) | 1 | 1 |
| **RF Frequency (GHz)** | 0.4 | 1.3 | 1.3 |
| **Repetition rate (Hz)** | 3067 | 5 | 3067 (CW) |
| **Avg. beam current (mA)** | 29 | 0.021 | >30 |
| **Particles per bunch $10^{10}$** | 15 | 2 | 2 |
| **Bunches per beam** | 328 | 1312 | CW |
| **Energy spread (rms, $10^{-3}$)** | 1.65 | 1.2; 0.7 | 1.2; 0.7 |
| **Beam transverse dimensions at IP (nm)** | 36 | 6 | 6 |
| **Beam $\sigma_z$ (rms, mm)** | 5.3 | 0.3 | 0.3 |

**Table 2. Parameters of dual axis asymmetric cavities of different configurations**

| Type of cavity | TT&LL + TT | LL & TT+LL |
|---|---|---|
| Operating frequency (GHz) | 1.300027 | 1.299977 |
| Frequency of the nearest mode (GHz) | 1.299311 | 1.29931 |
| $E_p/E_a$ | 2.71 | 2.85 |
| $B_p/E_a$ (mT/MV/m) | 6.1 | 6.35 |
| $R/Q$ (Ohm) - axis 1 | 380 | 381 |
| $R/Q$ (Ohm) - axis 2 | 289 | 291 |
| $V_z$ (MV) - axis 1 | 1.76 | 1.76 |
| $V_z$ (MV) - axis 2 | 1.53 | 1.54 |
| $R/Q$ (Ohm) - axis 1 | 203 | 211 |
| $R/Q$ (Ohm) - axis 2 | 281 | 296 |
| $G$ (Ohm) - operating mode | 276.8 | 279.2 |

**Table 3. Example of possible basic parameters of RERS and comparison with PERL and CBETA**

|  | RERS | PERL | CBETA |
|---|---|---|---|
| RF frequency (GHz) | 1.3 | 0.801 | 1.3 |
| Injection energy (MeV) | 40 | 5-10 | 6 |
| Repetition rate (MHz) | CW | CW | 325 |
| Particle Energy (GeV) | 1 | up to 1 | 0.15 |
| Beam current (mA) | 100-1000 | 10-20 | 40 (injector current) |
| Type of cavity | SRF dual axis | SRF single axis | SRF single axis |
| ΔE per cryo-module (MeV) | 80 MeV | 75 MeV | 36 MeV |
| Circumference (m) | <500 | <100 | 79.1 |
| Beam transverse dimensions $\sigma_\perp$ (μm) | 10-100 |  | 52-1800 |
| Beam $\sigma_z$ (ps/μm) | 0.1-1/30-300 | 10/3000 | 4/1200 |

**Table 4. Examples of possible basic parameters of the CLER-C storage ring (StRg).**

|  | CLER-C StRg | ILC Cooling Ring |
|---|---|---|
| Circumference (km) | ~6 | ~6 |
| RF frequency (GHz) | 1.3 GHz | 1.3 GHz |
| Energy (GeV) | 1 GeV | 5 GeV |
| Number of microbunches | CW | 1312 |
| $I_{av}$ (mA) | 100-1000 |  |
| Beam $\sigma_z$ μm | 30-300 | >300 |
| Application | Beam conditioning, Beam storage; photon factory | Beam storage, cooling and dumping. |

**Table 5. Comparison of possible LERC parameters and ILC parameters.**

|  | LERC | ILC |
|---|---|---|
| Type | ERL | LINAC |
| Cavity type | SRF | SRF |
| RF frequency (GHz) | 1.3 | 1.3 |
| Initial energy (GeV) | 1 | 15 |
| Final energy (GeV) | 250 | 250 |
| Average beam current (mA) | >30 | 0.021 |
| Beam transverse dimensions $\sigma_\perp$ at IP (nm) | 6 | 6 |
| Beam $\sigma_z$ μm (at IP) | 0.3 | 0.3 |
| Luminosity $\times 10^{34}$ cm$^{-2}$s$^{-1}$ | >10 | 1.35 |

**Figure captions**

**Figure 1.** Schematics of the possible configuration of the Circular-Linear Energy Recovery Collider showing the main stages of the system with: (a) the storage rings located at each end of the linear collider accelerating/decelerating arms; (b) the storage ring located centrally similar to the ILC designed

**Figure 2. (a)** The drawing of the dual-axis cavity with Tesla like cells along axis 1 (red dashed line) and the Low-Loss + Tesla cavity cells along the second axis. **(b)** Illustration of the operating mode structures along the both axes showing the possibility to tune the amplitudes of the operating mode by changing the cell's geometry.

**Figure 3.** Illustration of tunability and flexibility of the dual axis SRF structure architecture **(a)** shows the operating eigenmode amplitudes on both axes with higher amplitude for deceleration section to accept the spent beam and compensate for possible energy loss; (b) shows the system with a cathode implemented into the ERL structure.

**Figure 4. (a)** Schematic of the intermediate energy (1 GeV), large (2 km) diameter, weekly radiating (below 1kW) storage ring with indication of the FEL stations for photon production and research; **(b)** illustration of an example of the FEL station with dog-bone beam line for the beam energy recovery.

**Figure 5.** Examples of the beam injection into storage ring systems: **(a)** Recirculating Energy Recovery system; **(b)** conventional ERL based on dual axes asymmetric cavity.

**Figure 6.** The schematic of the Liner Energy Recovery Collider based on dual-axes asymmetric cavities.

**List of Tables**